\documentclass[aps,prl,preprint,groupedaddress]{revtex4-1}

\usepackage{graphicx}

\begin{document}


\title{Room temperature terahertz polariton emitter}


\author{Markus Geiser}
\email[]{mgeiser@ethz.ch}
\affiliation{Institute for Quantum Electronics, ETH Zurich, Wolfgang-
Pauli-Strasse 16, 8093 Zurich, Switzerland.}
\author{Giacomo Scalari}
\affiliation{Institute for Quantum Electronics, ETH Zurich, Wolfgang-
Pauli-Strasse 16, 8093 Zurich, Switzerland.}
\author{Fabrizio Castellano}
\affiliation{Institute for Quantum Electronics, ETH Zurich, Wolfgang-
Pauli-Strasse 16, 8093 Zurich, Switzerland.}
\author{Mattias Beck}
\affiliation{Institute for Quantum Electronics, ETH Zurich, Wolfgang-
Pauli-Strasse 16, 8093 Zurich, Switzerland.}
\author{J\'{e}r\^{o}me Faist}
\email[]{jerome.faist@phys.ethz.ch}
\affiliation{Institute for Quantum Electronics, ETH Zurich, Wolfgang-
Pauli-Strasse 16, 8093 Zurich, Switzerland.}


\date{\today}



\maketitle



{\bf The strong-coupling regime between an electronic transition and the photonic mode of a optical resonator~\cite{HAROCHE:1982p2071} manifests itself in the lifting of the degeneracy between the two modes and the creation of two polariton states with mixed optical and electronic character.  This phenomenon has been studied in atoms~\cite{Raimond:2001p2070}, excitons in semiconductors~\cite{Weisbuch:1992p12} and quantum electrodynamics circuits based on Josephson junctions ~\cite{Wallraff:2004p1760}. Recently, there is also strong interest to study similar effects using intersubband transitions in quantum wells~\cite{Dini:2003p1610} in the terahertz, where the  ultra strong coupling regime can be reached~\cite{Todorov:2010p1829} and new physical effects have been predicted~\cite{Ciuti:2005p1558}. An other interesting feature of this system is that, in contrast to systems based on superconductors, the ultra strong coupling regime can be maintained up to room temperature~\cite{Anappara:2009p1404, Geiser2010}. In this work, we demonstrate that parabolic quantum wells coupled to LC circuit resonators in the ultra strong coupling regime can achieve terahertz emission up to room temperature .} 


Spontaneous emission occurs through the coupling of a two-level system in the excited state with an available mode of the vacuum. For low frequencies  this process becomes less and less efficient as the density of available states decreases. This leads to the well-known $\lambda^2$ dependence of the spontaneous emission time with wavelength of the Einstein's spontaneous emission rate. The situation is especially dire in the terahertz where coherent emitters are also very difficult to realize~\cite{thz_review}. One possible way to circumvent this limitation has been to work in a system where the photonic density of state is enhanced by a small optical resonator, enhancing the  spontaneous emission by a Purcell factor ~\cite{Todorov:2007p705, Walther:2011p1891}.Although a Purcell factor up to 17 in an electrically pumped structure was reached ~\cite{Walther:2011p1891} both approaches did not allow the emission to be observed up to room temperature so far. 

Recently, a new approach was proposed where an electronic system is put in the strong coupling regime with the electrical field of an optical microresonator. As a result, a pair of excitations are created, the upper and lower polaritons, that are mixed states of the photon and electronic excitations at energies that are now shifted from the bare transition by an energy $ \pm \hbar \Omega_R$ where $\Omega_R$ is the vacuum Rabi frequency of the system. An especially interesting situation arises in the so-called ultrastrong coupling regime, reached when the ratio $\Omega_R / \omega$ of the vacuum Rabi frequency to the bare transition reaches the order of unity and, as the result, the ground state of the polaritonic system contains a non-zero average number of photons. Theoretical predictions in this case include, upon non-adiabatic modulation of the electron density, emission of correlated photon pairs analogous to a dynamical Casimir effect~\cite{ultrastrong1, ultrastrong2, DCE_Wilson}.

Electrical injection of polaritons is also very interesting for emission properties, as the polariton can be made to decay preferentially as a photon by designing the photon lifetime of the optical resonator to be shorter than the lifetime of the electronic excitation. As a result, the radiative efficiency of this quasi-particle can be made close to unity even in the presence of a short non-radiative lifetime. Enhancement of the radiative efficiencies from $10^{-7}$ to $10^{-1}$ together with the increase of the Rabi coupling energy $\Omega_R$ has been predicted theoretically~\cite{ciuti_emission_PRB} using a simplified rate-equation approach.

Polaritonic emission in the near infrared was recently observed at temperatures close to room temperature in a device based on an excitonic transition in a GaAs-based heterostructure~\cite{Tsintzos:2008p1895}. Electroluminescent polaritonic devices using intersubband-transitions have also been demonstrated in the mid infrared range \cite{Sapienza:2008p1406, el_2010} up too room temperature. In these devices, however, well-separated peaks could only be observed at low injection current as large injections reduced the polaritonic splitting. In addition, the geometry of the cavity enabled operation only in a narrow range of angles. 

In this work, we investigate the emission properties of small inductor-capacitor (LC) resonators operating in ultra strong coupling. These resonators enable a very tight confinement of the electric field to a volume on the order of $0.5\cdot10^{-3}\cdot\left(\lambda/n\right)^3$, greatly enhancing the vacuum electric field. As a material system we use intersubband transitions in modulation doped parabolic quantum wells. Their harmonic potential with equally spaced levels allows the absorption  to be insensitive to temperature, enabling the observation of strong coupling up to room temperature~\cite{Geiser2010}. In addition, thanks to Kohn's theorem, the position of the uncoupled resonance is not shifted by many-body effects~\cite{kohn,brey,kohn2011}. The polaritons are then excited by running an in-plane current in the quantum well system. 

As shown schematically and by a scanning electron micrograph in Fig. 1 a and b respectively, the sample consists of a field of LC resonators containing several hundred cavities with a pitch of 10~$\mu m$ and 20~$\mu m$ in the two normal directions on the surface on a $1mm^2$ area on top of the epitaxial material. As shown by a finite-element simulation in Fig.1 d, the electric field of the cavity modes is mainly oriented along the growth direction and therefore couples well to the intersubband transitions. A number of different samples were fabricated, with the resonator frequency lithographically tuned by scaling the electrical size of the central inductor part of the resonator. To allow in-plane electrical injection, the epitaxial material is separated from the metallic ground plane by a 0.12~$\mu$m thick Si$_3$N$_4$ layer. The sample is grown using molecular beam epitaxy and fabricated using epitaxial lift-off techniques. As shown schematically in Fig. 1 c, the active material consists of eight parabolic quantum wells, totaling a thickness of 0.88~$\mu$m.  The parabolic potential itself is formed by digitally alloying GaAs and Al$_{0.15}$Ga$_{0.85}$As layers ~\cite{Ulrich:1999p1550}. A sheet carrier density of $n_s = 3.2 \times 10^{11}$ is achieved by modulation doping the samples in thin 2nm thick GaAs wells in the center of the Al$_{0.3}$Ga$_{0.7}$As barriers separating the parabolic wells~\cite{kohn2011}. Finally, AuGe ohmic contacts are fabricated at the edges of the field to provide in-plane electrical excitation.

\begin{figure}
\includegraphics[width=0.5\textwidth]{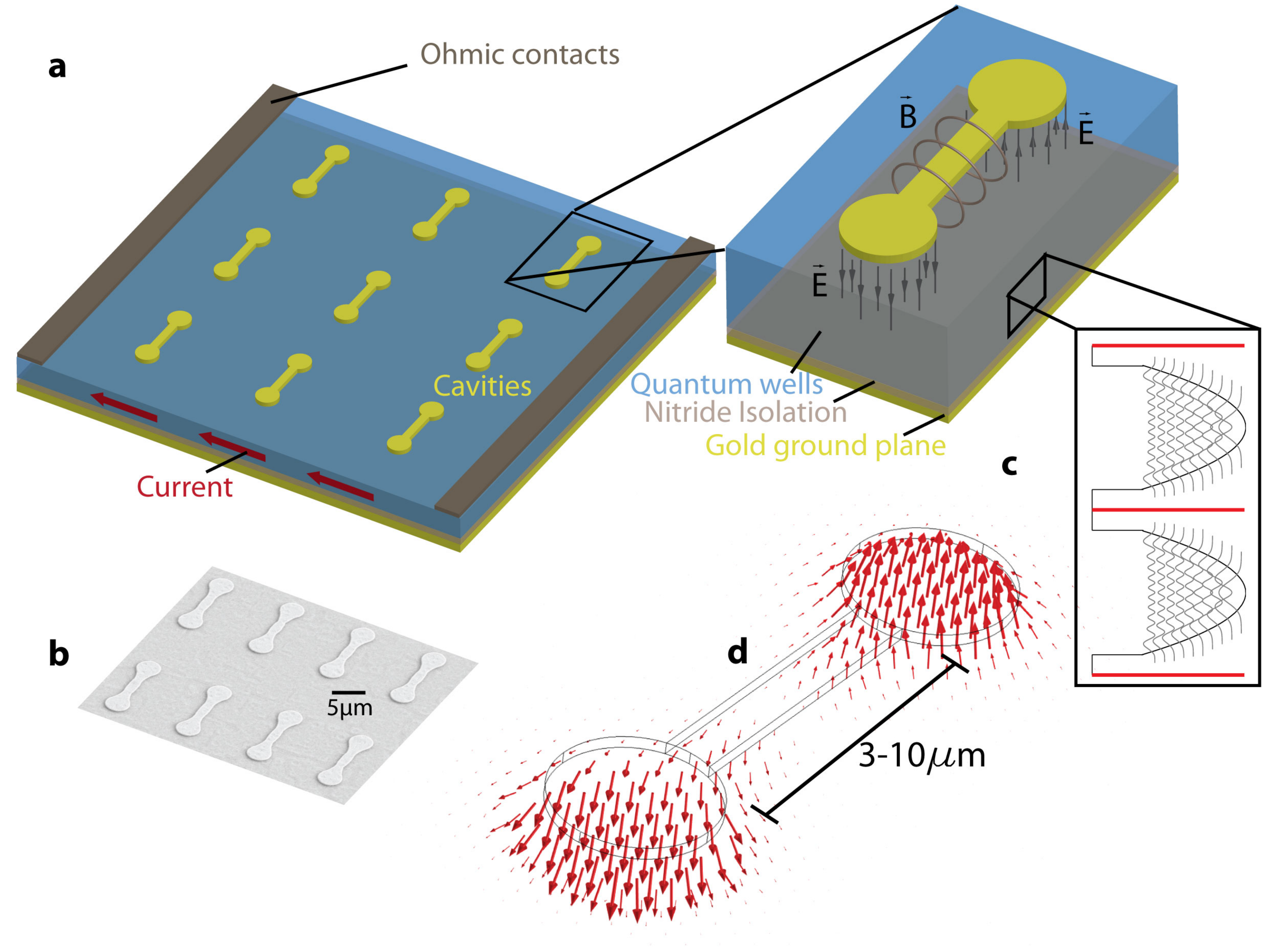}
\caption{\label{fig1} {\bf a} and {\bf b} display a schematic of a sample and a scanning electron micrograph respectively. A sample contains several hundred resonators on a $1$mm$\times1$mm area and Ge/Au ohmic contacts on two opposing sides of the sample for electrical contact. A zoom into the structure illustrates mode confinement in the cavity with an electric field oriented to couple well to the intersubband transitions. This behavior is confirmed by a finite element simulation shown in {\bf d}, where the arrows display electrical field orientation (arrow direction) and magnitude (arrow magnitude). The cavities are filled with eight parabolic quantum wells {\bf c}. Their modulation doping is indicated by the red lines.}
\end{figure}

Samples are indium soldered on copper heatsinks, wirebonded and mounted on the cold head of a helium flow crystat. The emission was measured by a home-made Fourier-transform infrared spectrometer (FTIR) operated under vacuum connected to a helium cooled silicon bolometer. The samples are electrically pumped with a $50$\% duty cycle to allow measurement in lock-in technique. The normalized emission spectra of samples with different cavity resonance frequencies at 10~K and 300~K are shown together in fig. 2 a and b. At 10~K, the spectra consist of two very clear emission peaks corresponding to the lower and upper polaritons that display a clear anticrossing behavior as a function of cavity resonance frequency. Close to the anticrossing point (third spectrum from the bottom in panel a, the lineshapes are similar with a linewdith of about 1.3meV showing the mixed character of the light and matter modes. 

\begin{figure}
\includegraphics[width=0.47\textwidth]{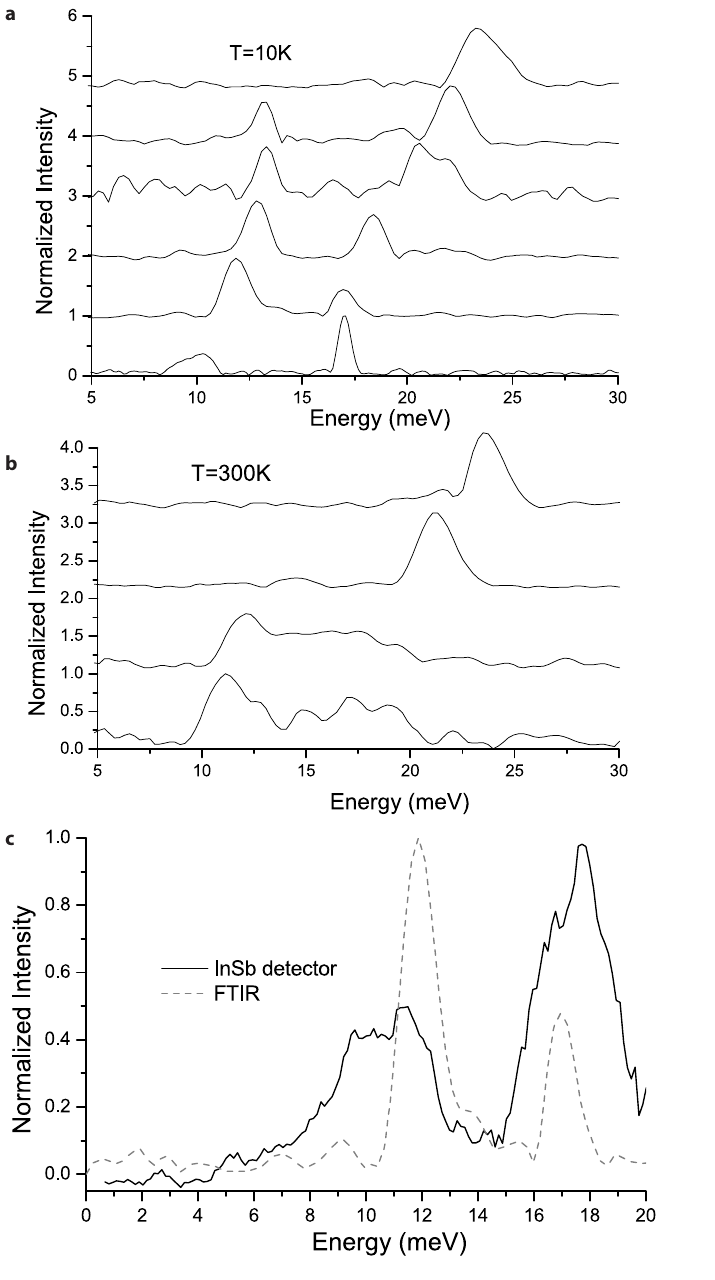}
\caption{\label{fig2}Normalized electroluminescence spectra with different cavity resonance frequencies at cryogenic a and room b temperature. At cryogenic temperatures, the spectra show two distinct peaks corresponding to the two polariton states. The peaks shift with cavity resonance and a clear anticrossing is observed. Close to the anticrossing point (third spectrum from the bottom), the quality factors and absorption strength of the peaks are very similar, owing to the mixed light matter character of the states. b At room temperature, the Rabi frequency of $\Omega_R=720GHz$  does not change but the emitted power drops. c Spectrum close to the anticrossing point measured in completely cold (4K) environment using a tunable InSb detector, compared to the one measured with the FTIR.}
\end{figure}

Room temperature spectra are shown in Fig. 2~b. The resonances are broader, but remain at the same frequencies compared to lower temperatures. The emitted power is  estimated to be on the order of 2~pW.
To prove that the power was indeed emitted from the sample and is not the result of a modulation of reflection of outside thermal radiation, we remeasured the low-temperature spectrum with an in-situ tunable InSb detector~\cite{Blaser:2002p277}. In the this experimental setup, both sample and detector share the same helium bath and all the external blackbody emission is screened by walls cooled at 4K. The corresponding spectrum is shown in fig. \ref{fig2}~c and exhibits the same double peak features as the one obtained by the FTIR.  

\begin{figure}
\includegraphics[width=0.47\textwidth]{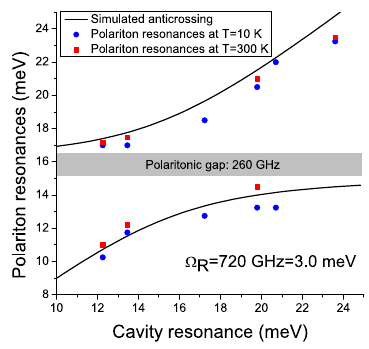}
\caption{\label{fig3} Position of the  emission peaks at cryogenic temperature (blue dots)  and room temperature (red squares), along with the predictions of a theoretical model (continous lines). Both data sets agree well with the model and reproduce a polaritonic gap of $260$~GHz.}
\end{figure}

When plotted against the bare cavity resonance (fig \ref {fig3}), the positions of the upper and lower polariton show clear anticrossing behavior, in good agreement with the predictions from a theoretical model~\cite{Todorov:2010p1829} treating the light-matter interaction in the dipolar gauge~\cite{kohn2011}. The presence of the insulator layer has been explicitly included in the electromagnetic modeling of the cavity. The large measured interaction strength, expressed in the Rabi frequency is $\Omega_R=720GHz \left( \widehat= 3.0meV\right)$, which is $18\%$ of the intersubband energy and does not change appreciably with temperature. A signature of the ultra strong coupling regime is the appearance of a 260~GHz polaritonic gap \cite{Todorov:2010p1829} accounting for a significant portion of the interaction energy.

In contrast to previous emission experiments that used planar Fabry-Perot microcavities near the total internal reflection angle~\cite{Sapienza:2008p1406}, in our dispersionless LC resonators the strong coupling is achieved for all incident angles. As a result, radiative efficiency can be thought a result of global lifetime modifications. 
A typical sample has up to $60$pW emitted power corrected for collection efficiency at $100$mW injected electrical power, corresponding to an conversion efficiency of $6 \cdot 10^{-10}$. The emitted power can be written as a product of efficiencies
\begin{equation}
P_{opt} = \eta_{geom} \frac{Q}{Q_{rad}} \frac{\tau_{p}^{-1}}{\tau_p^{-1} + \tau_{nr}^{-1}} \frac{2}{N}  P_{elec}
\label{efficiency}
\end{equation}
where the $\eta_{geom} = 2.75 \times 10^{-2}$ is a geometrical factor, product of the optical collection efficiency and fraction of the pumped area under the resonator. Using a 3D finite element simulation of the structure, the outcoupling efficiency of the cavity $Q/Q_{rad} = 0.014$ can be inferred from a radiative $Q_{rad} = 700$, and the global quality factor of the cavity $Q = 14$, which is dominated by Ohmic losses. The branching ratio $ \frac{\tau_{p}^{-1}}{\tau_p^{-1} + \tau_{nr}^{-1}} = 0.24$ is the ratio of the polaritons decaying into photon as compared to the electronic excitations (a nonradiative lifetime of $10$ps was used for low temperatures). This ratio remains favorable up to room temperature where the lifetime is expected to still be on the orders of ps. Finally the smallest number is the polariton injection efficiency, that is the ratio of the "bright" to "dark" excitations that are generated. As shown in ~\cite{DeLiberato:2009p1405}, in the polariton picture of interactions, two bright excitations (the upper and lower polaritons) are created along with N dark excitations. As we are exciting these polaritons thermally and assuming a high temperature limit ($kT \geq \hbar \Omega_{R}$) we then can assume $\eta_{polariton} = 2/N = 5.1 \times 10^{-6}$ where N is the number of electrons in quantum wells inside the LC resonator. The benefit of a small cavity is immediatly apparent as N is proportional to the effective volume of the cavity. The product of these efficiencies is $4.7 \times 10^{-10}$ is reasonably close to the measured value of $6 \times 10^{-10}$.

As a result of Eq.~\ref{efficiency}, we expect better temperature performance of the polariton electroluminescence compared to an an intersubband electroluminescent diode. In fact, the key difference is that the much longer spontaneous lifetime $\tau_{spon}$ replaces $\tau_{p}$ in the expression for the branching ratio. The device demonstrated here is only a proof of principle. By injecting the current only under the resonators and choosing a resonator geometry with a much lower radiative Q, improvement of the efficiency by many order of magnitude should be obtained, enabling a device with an optical power in the microwatt range at room temperature.

This work was supported by the Swiss National Science Foundation through contract no $200020\_129823/1$.

\end{document}